\documentclass[aps,prl,twocolumn,superscriptaddress,nofootinbib,showpacs,showkeys,preprintnumbers]{revtex4-1}
\usepackage{graphicx}  % needed for figures
\usepackage{dcolumn}   % needed for some tables
\usepackage{bm}        % for math
\usepackage{amssymb}   % for math
\usepackage{lineno}
\usepackage{draftcopy}

\usepackage{color}

\usepackage[bookmarks]{hyperref}

\def \deta {\Delta \eta} 
 
\newcommand{ \be }{\begin{eqnarray}}
\newcommand{ \ee }{\end{eqnarray}}

\newcommand{ \eps }{\varepsilon}
\newcommand{ \mean }[1]{\left\langle #1 \right\rangle}   

\newcommand{ \psin }{\Psi_{n}}
\newcommand{ \psirp }{\Psi_{RP}}
\newcommand{ \phia }{\phi_{\alpha}}
\newcommand{ \phib }{\phi_{\beta}}

\begin{document}

%\setlength{\linenumbersep}{6pt}
%\linenumbers

%\widetext

\title{Ultra-relativistic nuclear collisions: event shape engineering}

\author{J\"urgen Schukraft}
\affiliation{PH Division, CERN, CH-1211 Geneva 23, Switzerland}
\author{Anthony Timmins}
\affiliation{University of Houston, Houston, TX 77204 }
\author{Sergei A. Voloshin}
\affiliation{Wayne State University, 666 W. Hancock, Detroit, MI 48201}

\begin{abstract}
The evolution of the system created in a high energy nuclear collision
is very sensitive to the fluctuations in the initial geometry of the
system. In this letter we show how one can utilize these large
fluctuations to select events corresponding to a specific initial
shape. Such an ``event shape engineering'' opens many new
possibilities in quantitative test of the theory of high energy  nuclear
collisions and understanding the properties of high density hot QCD
matter.
\end{abstract}

\maketitle

Many features of multiparticle production in ultra-relativistic
nuclear collisions reflect the initial collision geometry of the system.  As the
initial conditions affect to a different degree all the particles, it
leads to truly multiparticle effects often referred to as anisotropic
collective flow.  Studying anisotropic flow in nuclear collisions
provides unique and invaluable information about the evolution of the
system created in a collision, properties of high density hot QCD
matter, and the physics of multiparticle production in
general~\cite{Voloshin:2008dg,Muller:2012zq}.  Recently, a significant
progress has been reached in understanding the role of the
fluctuations in the initial density
distribution~\cite{Mishra:2007tw,Sorensen:2009cz,Takahashi:2009na,Alver:2010gr,Teaney:2010vd}.
In particular it was realized that such fluctuations lead to odd
harmonic anisotropic flow, which enable new insights into
dynamics of the system evolution. The experimental
measurements~\cite{Aamodt:2010pa,ALICE:2011ab} confirm the existence
of collective flow up to at least sixth harmonic, thus validating the
picture.

At present, the effect of the initial geometry on final state
observables can be studied only by varying the collision centrality,
or colliding nuclei of different size and shape.  It has
been always tempting to study anisotropic flow at maximum particle
density, but this was possible only in very central collisions where
the anisotropies are small.  Collisions of very non-spherical nuclei,
such as uranium, should be able to provide events with large initial
anisotropy and high particle density (in the so-called body-body
collisions), however the analysis might be very complicated due to
large variety of possible overlap geometries that have to be
experimentally disentangled.  In this paper we discuss how one can
select events corresponding to different initial system shapes
utilizing strong fluctuations in the initial geometry even at fixed
impact parameter,  e.g. Au+Au central collisions but of large initial
anisotropy, and in this way to study the system evolution under
conditions not possible before.

The study of particle production in the events corresponding to a
specific geometry opens a number of very attractive possibilities.
One of those, mentioned above, is the study of the system evolution in
a high density regime (central collisions) and concurrently strongly
anisotropic initial conditions. This would add new constraints to
questions such as the approaching of the system evolution toward the
so-called ``hydrodynamic limit'' and the development of the
anisotropic flow velocities fields.  Analysis of transverse momentum
spectra in such events can shed light on correlation between radial
and anisotropic flow.  Another example would be understanding the
``away-side'' double bump structure in two-particle azimuthal
correlations~\cite{Adams:2005ph,Adare:2008ae}.  Several years ago, this
attracted a considerable attention as a possible indication of the
Mach cone due to propagation of a very energetic parton through the
dense medium. More recently it was found that this structure is likely
due to triangular (third harmonic) flow. Additional proof for this
interpretation might come from studying such correlations in events
with very small triangularity. Several other examples,
including azimuthally sensitive femtoscopy and an estimate of the
background effects in chiral magnetic effect studies will be discussed
later in the paper.

There might be different approaches to perform such an event shape
engineering (ESE).  The one adopted in this paper is an an extension
of the technique proposed in~\cite{Voloshin:2010ut} that is based on
the event selection according to the magnitude of the so-called
reduced flow vector $q_n$ (the subscript $n$ is the harmonic number,
for the exact definitions see below).  We always perform ESE using two
subevents. We use here a common terminology in flow analyses, where a
subevent refers to a distinct subset of all measured particles
selected either at random or in a given momentum region.  One of the
subevents is used for the event selection (we will always call it
subevent ``a'' below) whereas the physical analysis is performed on
the second subevent (subevent ``b'').  Using two subevents helps to
avoid nonphysical biases due to nonflow effects as discussed below.
We use Monte-Carlo Glauber model to illustrate how the event selection
based on flow vectors works and outline the general scheme for the
corresponding experimental analysis. 

To quantify the anisotropic flow we use a standard Fourier
decomposition of the azimuthal particle distribution with respect to
the $n$-th harmonic symmetry
planes~\cite{Voloshin:1994mz,Poskanzer:1998yz}:
\begin{equation}
E\frac{d^3 N}{d^3 p} = \frac{1}{2\pi} \frac{d^2 N}{p_{T} dp_{T}
  dy}\! \left(1\!+\!\sum_{n=1}^{\infty}2v_n \cos[n\!\left(
\phi\!-\!\Psi_n \right) ] \!\right), 
\label{eqFourier}
\end{equation}
where $v_n$ is the $n$-th harmonic flow coefficient and $\psin$ is
the $n$-th harmonic symmetry plane determined by the initial geometry
of the system (as given by the participant nucleon distribution, see below).
The event-by-event fluctuations in anisotropic flow are believed to
follow the fluctuations in the corresponding eccentricities of the
initial density distribution.  Following~\cite{Teaney:2010vd}, for the
latter we use the definition
\be
\eps_{n,x}=\mean{r^n \cos(n \phi)},\; \eps_{n,y}=\mean{r^n \sin(n
  \phi)} \\ \eps_{n,p}=\sqrt{\eps_{n,x}^2+\eps_{n,y}^2}, \;\;\;
\tan(n\Psi_n)=\eps_{n,y}/\eps_{n,x},
\ee 
where $\eps_{n,p}$ is the so-called {\em participant} eccentricity.
The average can be taken with energy or entropy density as a weight.
In our Monte-Carlo model we weight with the number of participating
(undergoing inelastic collision) nucleons.  For the nucleon
distribution in the nuclei we use Woods-Saxon density distribution
with the standard parameters (for the exact values
see~\cite{Voloshin:2010ut}); the inelastic nucleon-nucleon cross
section is taken to be 64~mb.  We assume that the flow values are
proportional to the corresponding eccentricities with the ratio fixed
to approximately reproduce measured $v_n$ values~\cite{ALICE:2011ab}.
As it is shown in~\cite{Voloshin:2007pc}, in this case the
distribution in $v_n$ is very well described by the so-called
Bessel-Gaussian (BG) distribution ${\rm BG}(v;v_0,\sigma_{vx})$, where
\be {\rm BG}(x;x_0,\sigma) =
\frac{x}{\sigma} I_0\left(\frac{x_0\, x}{\sigma^2}\right)
\exp\left(-\frac{x_0^2+x^2}{2\sigma^2}\right), 
\ee 
which is a radial projection of 2-dimensional Gaussian distribution
with the width $\sigma$ in each dimension and shifted off the
  origin by distance $x_0$.

The flow vectors are calculated in two
subevents~\cite{Poskanzer:1998yz,Voloshin:2008dg} with multiplicities
in each subevent approximately corresponding to $\deta=0.8$ in
Pb+Pb collisions at LHC energies~\cite{Aamodt:2010cz}
(approximately 1200 charged particles per subevent for 0--5\%
centrality).  The multiplicities are generated with a negative
binomial distribution based on number of participants and number of
the binary collision as in~\cite{Voloshin:2010ut}.  For each 5\% width
centrality bin discussed below, we analyze about 1.2~M simulated
events.

The flow vectors are defined as
\be
&&Q_{n,x}=\sum_i^M \cos(n\phi_i);\;   
Q_{n,y}=\sum_i^M \sin(n\phi_i);\;
\\
&&q_n=Q_n/\sqrt{M},  \;
%\nonumber
\label{eq:q} 
\\
&&
q_n^2=1+(M-1)\mean{\cos[n(\phi_i-\phi_j)]}_{i\ne j}
%\nonumber
\label{eq:qV} 
\ee
where $M$ is the particle multiplicity and $\phi_i$ are the particle
azimuthal angles of particles in a given subevent.  Eq.~\ref{eq:qV}
presents the relation of the length of $q_n$ vector to the average
correlation between all pairs of particles in a given event.  The
event-by-event distribution in the magnitude of flow vectors $q_n$ has
been proposed~\cite{Voloshin:1994mz} and often used to measure the
average flow~\cite{Adams:2004bi,Voloshin:2008dg}.  The distribution in
$q_n$ is determined by the $v_n$ distribution convoluted with
statistical fluctuations due to finite multiplicity. For relatively
high multiplicities ($M \gtrsim 300$) it is very well described by BG
distribution ${\rm BG}(q;q_0,\sigma_{qx})$ with parameters related to
those of $v_n$ distribution: \be q_0=\sqrt{M} \,v_0,\;\;
\sigma^2_{qx}=\frac{1}{2}\left[1+(M-1)(2\sigma^2_{vx}+\delta)\right],
\ee where $M$ is the multiplicity used to build the flow vector, and a
nonflow parameter $\delta$ accounts for possible correlations not
related to the initial geometry of the system. (For a more detailed
discussion of the functional form of $q_n$ distributions
see~\cite{Sorensen:2008zk}.) Thus, the fit to $q_n$-distribution
provides information about underlying flow fluctuations, {\it if} the
nonflow contribution can be neglected or estimated from other
measurements.

\begin{figure}
\includegraphics[keepaspectratio, width=1.\columnwidth]{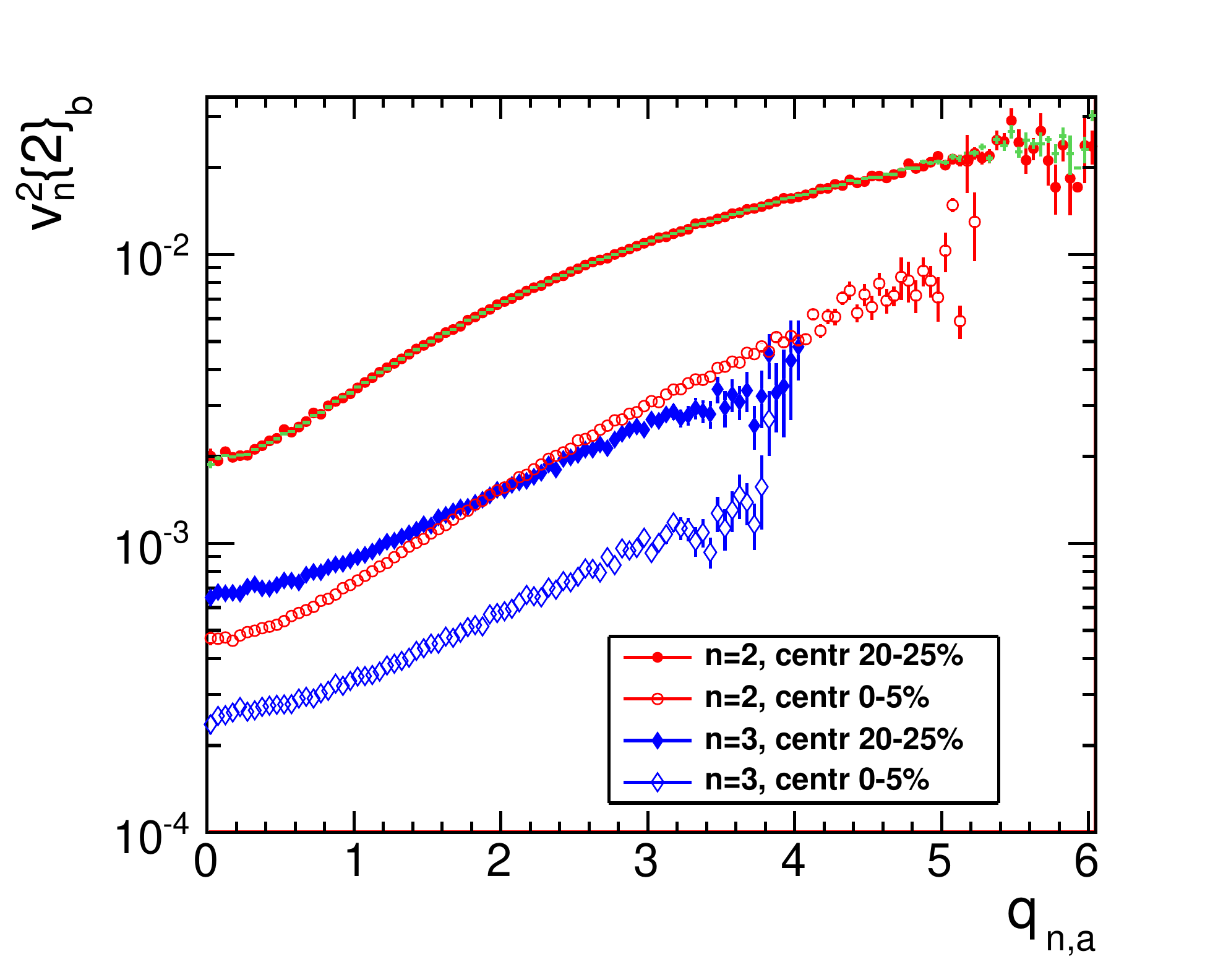}
  \caption{(color online) Mean elliptic and
    triangular flow values in $a$-subevent as function 
    of the corresponding $q_n$  magnitude in $b$-subevent. }
  \label{fig:vnq}
\end{figure}

\vspace*{1mm}
\noindent {\it Zero nonflow.} We start the discussion of the ESE with
the simplest case when all the correlations in the system are
determined only by anisotropic flow.  Figure~\ref{fig:vnq} shows the
average values of $v_n^2$ calculated via 2-particle correlation method
in one of the subevent (``b'') as function of the flow vector
magnitude in the second subevent (``a'').  We remind the reader, that
in this simulations the two subevents are statistically
independent and are correlated only via common participant plane and
flow values. There are no nonflow correlations
included at this stage.  In this case the results for $v^2_{n,b}\{2\}$
coincide with ``true'' values of $\mean{v^2_n}$ (not shown), though
have slightly larger statistical errors due to finite multiplicity of
the subevent.

The results in Fig.~\ref{fig:vnq} demonstrate, that depending
on $q_{2,a}$ one can select events with average flow values varying
more than a factor of two.  How well one can ``resolve'' the flow
fluctuations depends on the number of particles used to calculate the
flow vector as well as, though weakly, on flow magnitude itself. 
We find that for centrality 20--25\%, the width of the $v_2$
distribution for a fixed $q_{2,a}$
value is about factor of 1.5 smaller than that for unbiased event
sample (changing from 0.031 to 0.022);
it decreases for about 20\% if one double the size of the
subevent (double the multiplicity) used for $q_2$ determination. 

\begin{figure}
\includegraphics[keepaspectratio, width=1.\columnwidth]{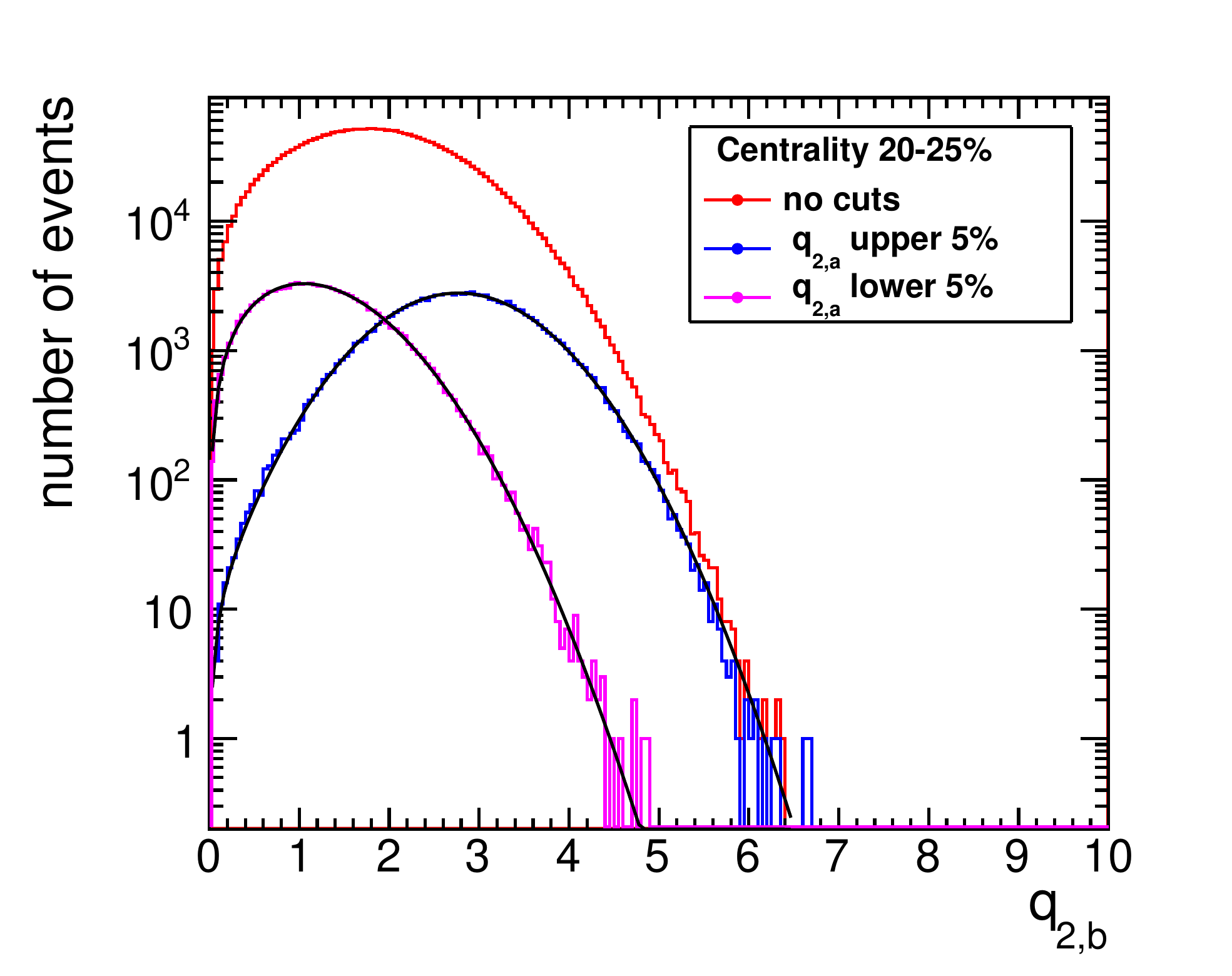}
\includegraphics[keepaspectratio, width=1.\columnwidth]{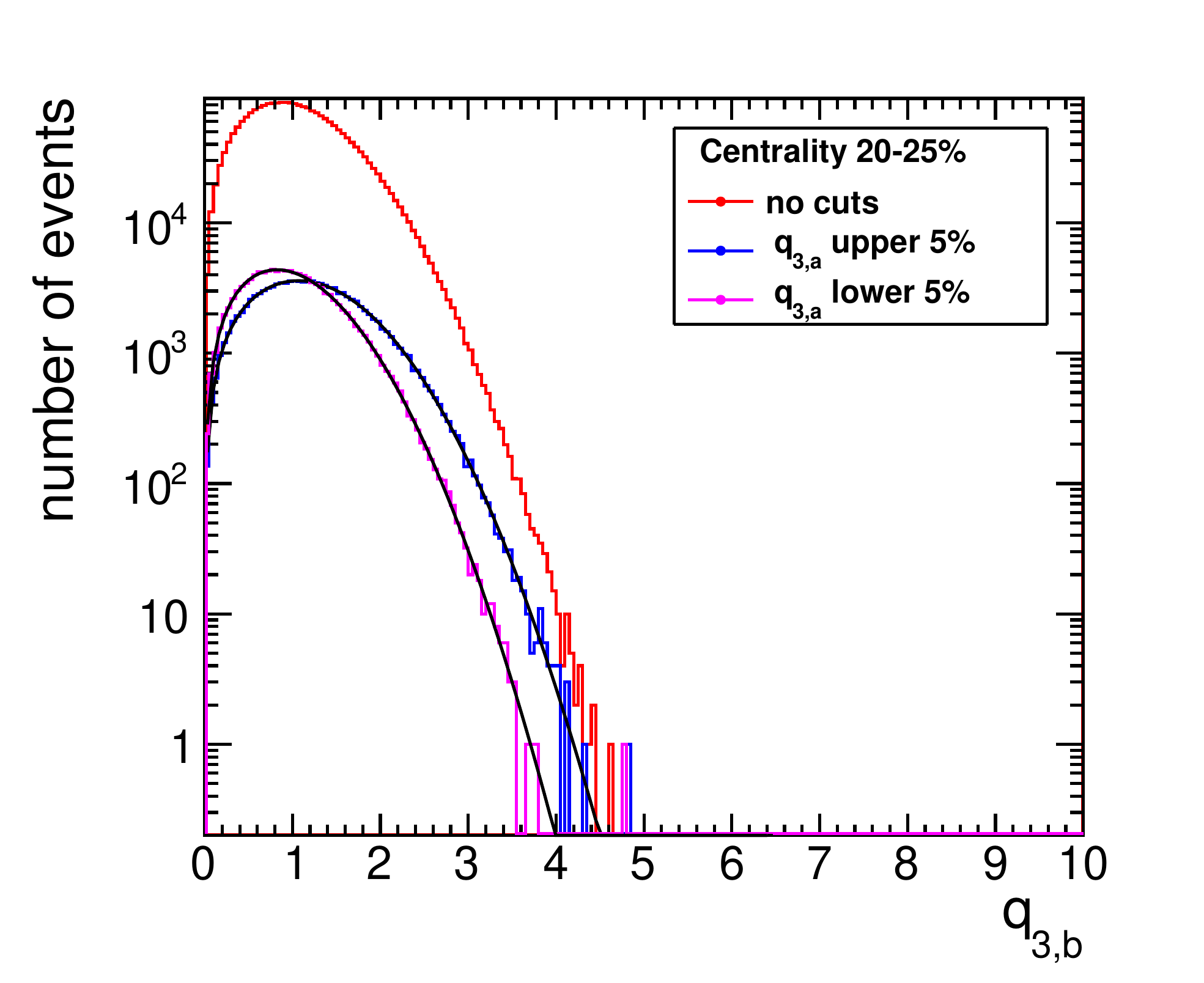}
  \caption{(color online)
    $q_{2,b}$ and $q_{3,b}$ distributions in the event samples selected by
    different cuts on the corresponding 
    $q_{n,a}$-vector magnitude indicated in the plot. The lines show
    the BG fit to the distribution.}
  \label{fig:qn}
\end{figure}

\begin{figure}
\includegraphics[keepaspectratio, width=1.\columnwidth]{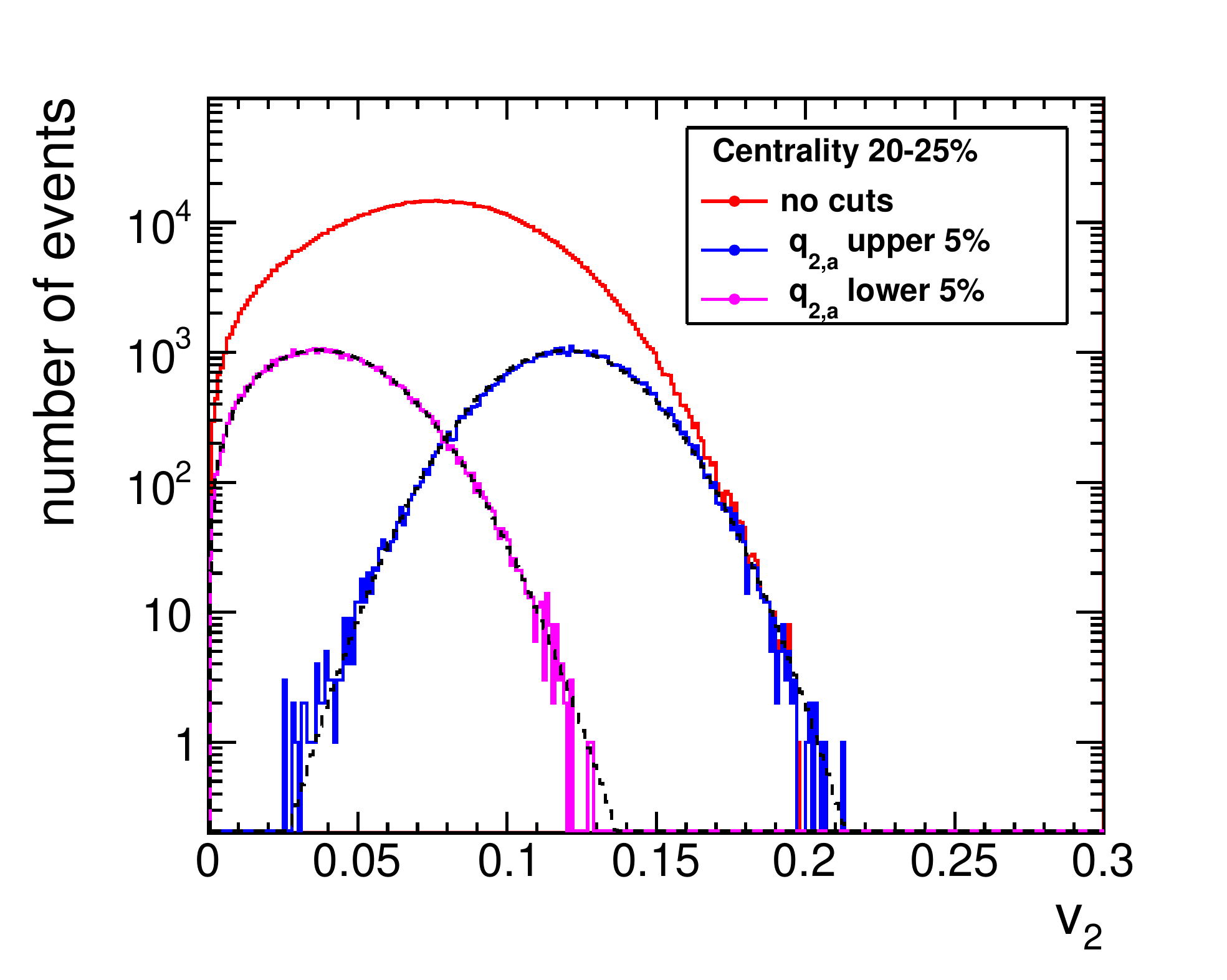}
\includegraphics[keepaspectratio, width=1.\columnwidth]{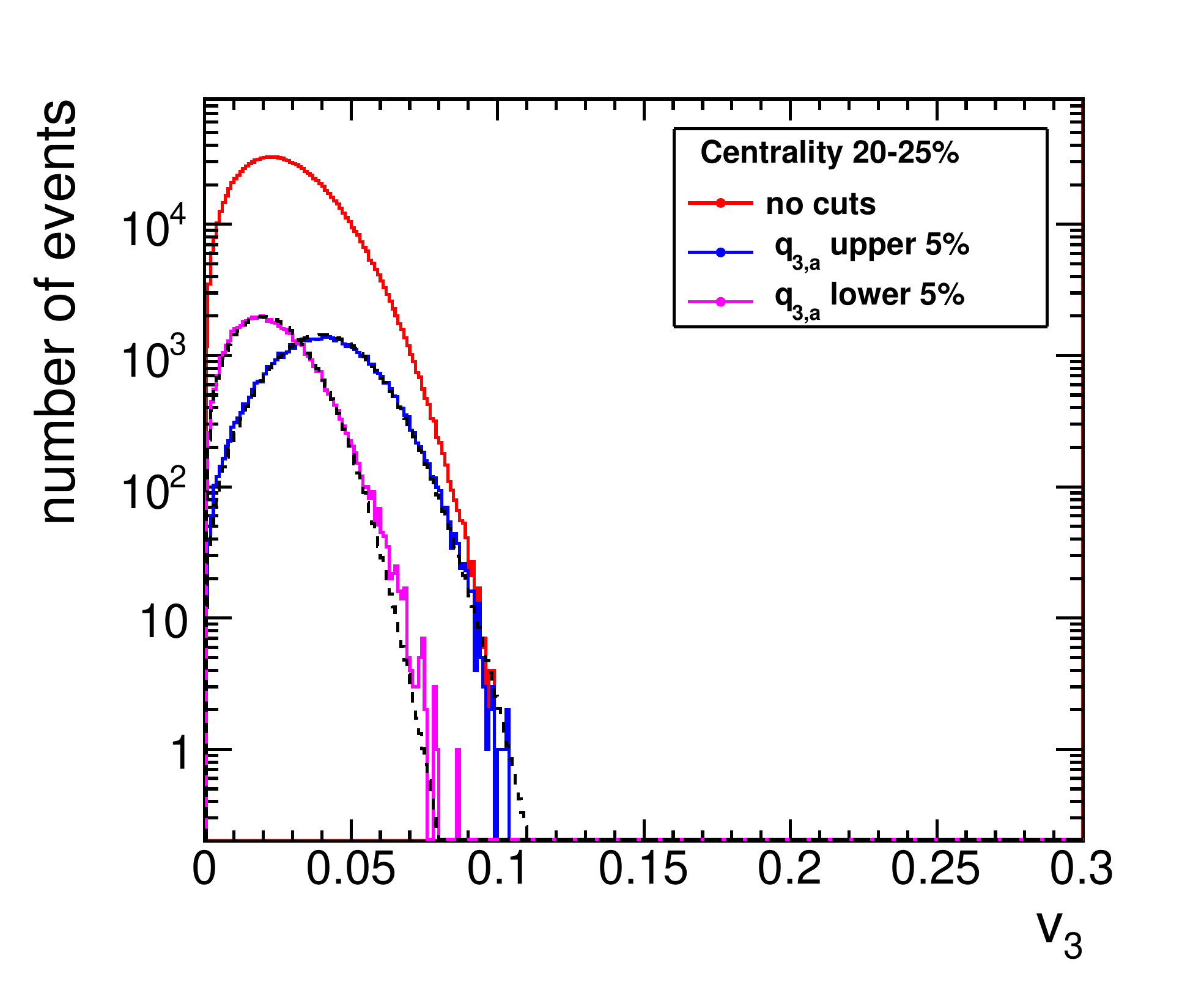}
  \caption{(color online) Actual (true) $v_2$ and $v_3$ distributions
    in the event samples selected by different cuts on the
    corresponding $q_n$-vector magnitude indicated in the plot
    compared to that extracted from the BG fits to $q_{n,b}$
    distribution shown in Fig.~\ref{fig:qn} (dashed lines).  Note that
    the lines are not the fit to the histograms!  }
  \label{fig:vn}
\end{figure}

Let us demostrate now how in practice one can obtain an information
about the $v_n$ distributions, corresponding to different cuts on
the $q_{n,a}$ values, from the fits to the $q_{n,b}$-distributions.
Figure~\ref{fig:qn} shows distribution in $q_{n,b}$ (subevent-b) for
three different cuts on $q_{n,a}$, separately for the second and third
harmonic flow.
All $q_{n,b}$ distributions in Fig.~\ref{fig:qn} are fit to the
BG functional form to extract the corresponding mean
flow values and the corresponding width (see,
e.g. ~\cite{Voloshin:2008dg}). It is remarkable that the fits are very
good not only for the unbiased $q$-distributions but also to the ones
corresponding to the low flow and high flow ``engineered events''
(corresponding to the 5\% lowest and 5\% highest $q_{n,a}$ events).
Using the extracted parameters we plot the corresponding $v_n$
distributions in Fig.~\ref{fig:vn} (shown by dashed lines) and compare
to the actual (``true'') $v_n$ distributions, which is known in this
Monte-Carlo simulation (shown as a histogram).  One finds an excellent
agreement between the two indicating that the $v_n$ distributions in the
``shape engineered'' events are very close to the BG form.

\vspace{1mm}
\noindent {\it Nonflow effects.}  The ESE approach described above is
based on using two subevents.  In this case possible
nonflow effects can be separated in two major categories (a) when
nonflow effects are present within each of the subevents, but there is
no nonflow correlations between subevents ``a'' and ``b'', and (b)
when nonflow correlations are present between, as well as within,
subevents.  As we show below one should try to minimize the nonflow
correlations between the two subevents which are used for ESE
selection and physical analisys, respectively.  A practical solution
to that might be to use subevents which are separated by a significant
(pseudo)rapidity gap.

The case (a) does present a certain challenges to the analysis, but no
more than the one in the conventional flow analysis. Once the event
selection is done with $q_{n,a}$ cuts, the flow in the selected events
can be estimated using particles in subevent ``b'' with standard
methods including many-particle cumulant analysis.  The case (b) is
significantly more complicated. Below we only discuss possible biases,
without trying to resolve the problem.

We simulate nonflow effect by assuming that half of all particles in
the entire event are produced in pairs with both particle in a pair
emitted with the same azimuthal angle. Each particle is assigned
randomly to one of the two subevents. In this case the nonflow
parameter $\delta=1/(2M)$, where $M$ is the (full) event multiplicity,
which roughly corresponds to the nonflow estimates in real LHC events
for particles at midrapidity $|\eta|<0.8$.

\begin{figure}
\includegraphics[keepaspectratio, width=1.05\columnwidth]{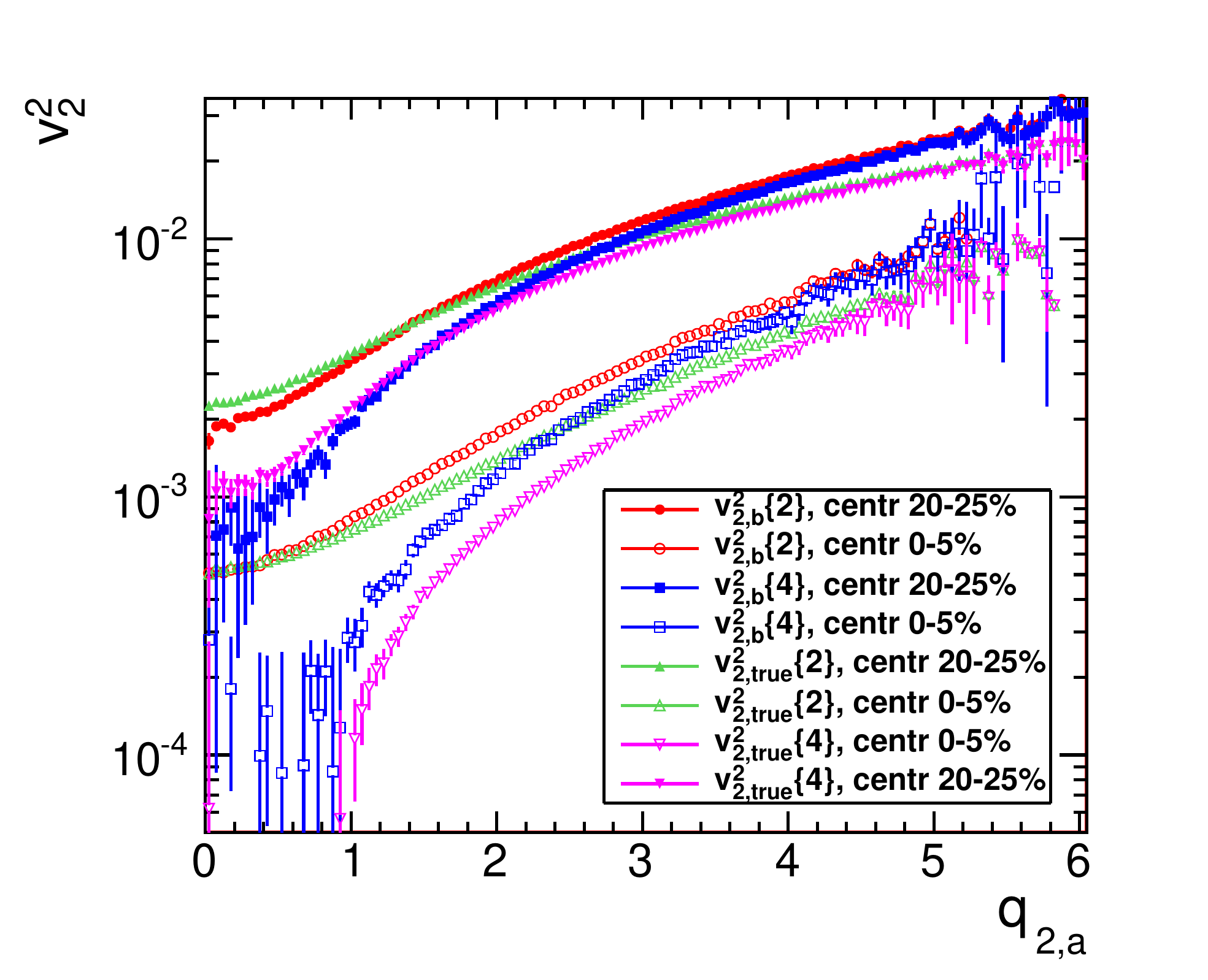}
  \caption{(color online) Elliptic flow measured with 2- (red points)
    and 4-particle (blue) cumulant method in subevent ``a'' as a
    function of the corresponding $q_{2,b}$ magnitude. Solid symbols
    correspond to centrality 20-25\%, and open symbols to 0-5\%
    centrality.  The true (simulated) values are shown by green
    markers, as expected for 2-particle cumulant results and by
    magenta for 4-particle cumulant results.}
  \label{fig:vnnf}
\end{figure}

\begin{figure}
\includegraphics[keepaspectratio, width=1.\columnwidth]{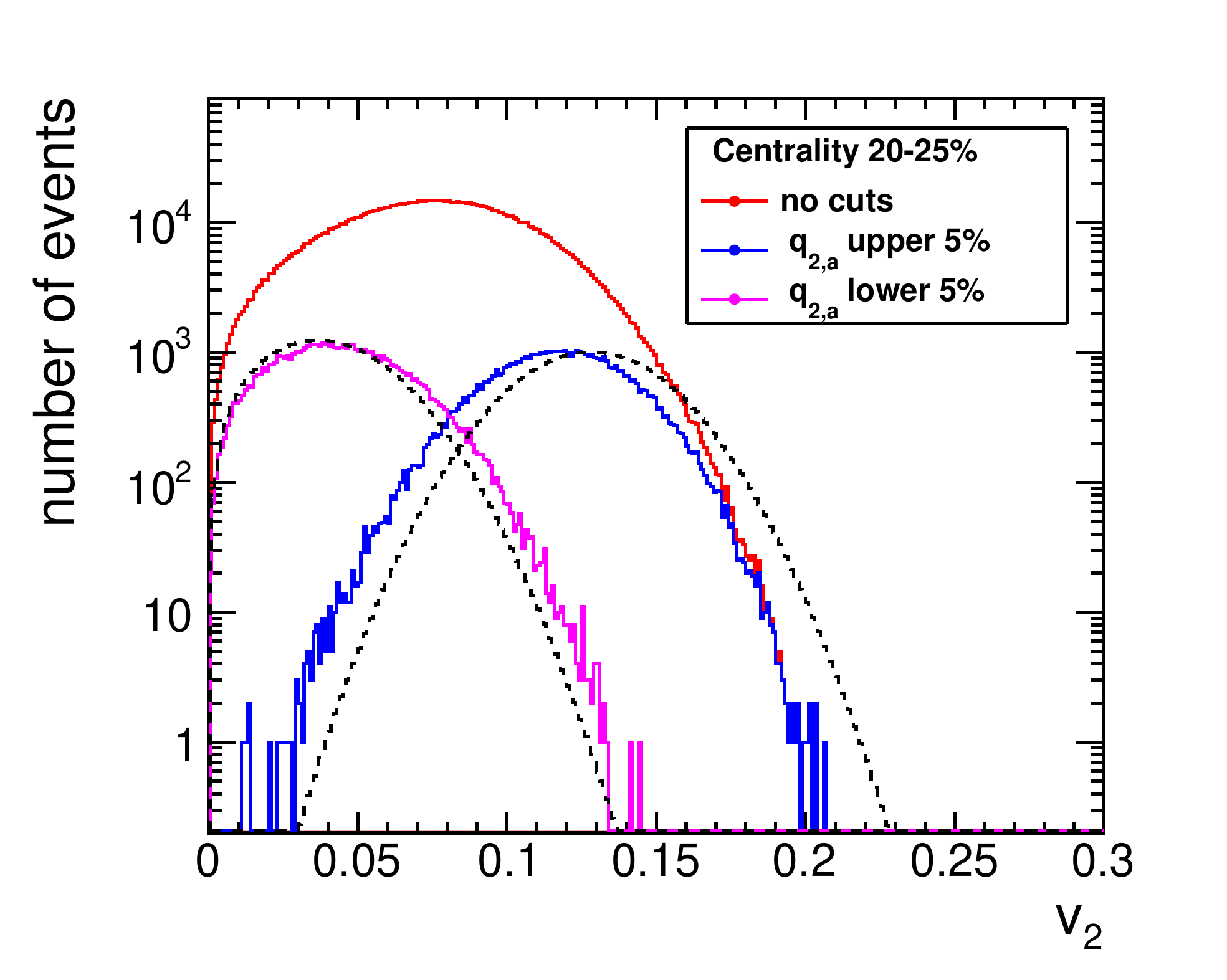}
\includegraphics[keepaspectratio, width=1.\columnwidth]{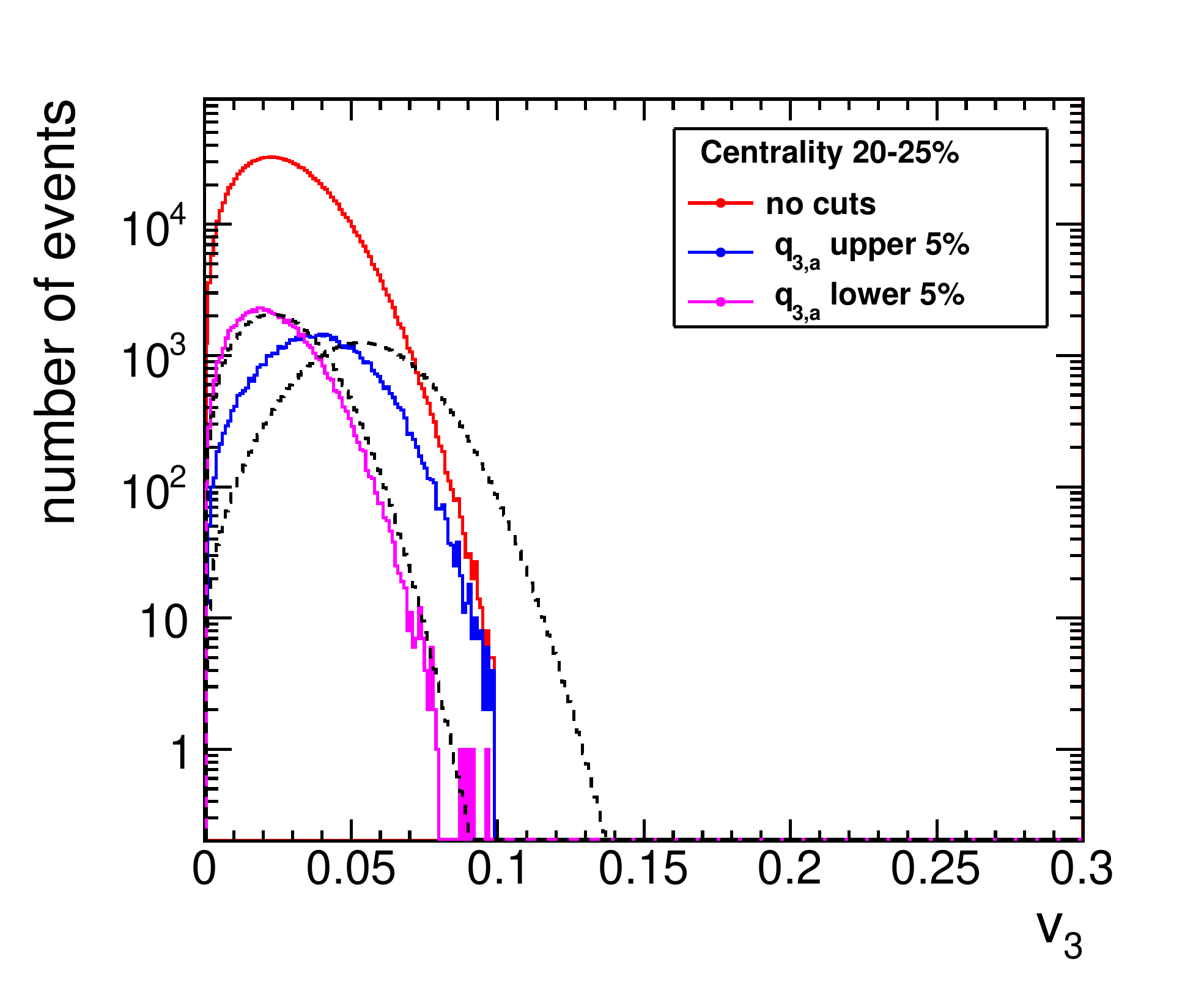}
  \caption{(color online)
    The same as in Fig~\ref{fig:vn} but for the case of nonflow
    described in text.}
  \label{fig:vn_nf}
\end{figure}

Figure~\ref{fig:vnnf} presents the results for flow calculation in
subevent ``a'' using 2- and 4-particle cumulant methods as function of
$q_{2,b}$. The expectations based on simulated flow are also
shown. One observes a significant bias due to nonflow, leading to
overestimate the flow values in high flow selected events and
underestimate in the low flow selected events. This trend is due to
positive character of the nonflow correlations.  The corresponding
bias in corresponding $v$ distributions is shown in
Fig.~\ref{fig:vn_nf}.
Note that even though the bias for mean values of flow is somewhat
modest, at large values of $v_n$ the actual distribution could differ
by order of magnitude from the one deduced from $q$-distribution fits.

Below we discuss very briefly several analyses, which can profit from
the event shape engineering.

\vspace{1mm}
\noindent{\em The chiral magnetic effect} proposed
in~\cite{Kharzeev:2004ey,Kharzeev:2007tn,Kharzeev:2007jp} is a charge
separation along the magnetic field.  A correlator sensitive to the
CME was proposed in Ref.~\cite{Voloshin:2004vk}:
\be
\hspace*{-2cm}
 \mean{ \cos(\phia +\phib -2\psirp) }, 
\label{eq:obs1}
\ee
where subscripts $\alpha,\; \beta$ denotes the particle type.  The
STAR~\cite{:2009uh,:2009txa}, as well as the
ALICE~\cite{Abelev:2012pa} collaboration measurements of this
correlator are consistent with the expectation for the CME and can be
considered as evidence of the local strong parity violation.  The
ambiguity in the interpretation of experimental results comes from a
possible background of (the reaction plane dependent) correlations not
related to CME.  Note that a key ingredient to CME is the strong
magnetic field, while all the background effects originate in the
elliptic flow~\cite{Voloshin:2004vk}. This can be used for a possible
experimental resolution of the question. One possibility is to study
the effect in central collisions of non-spherical uranium
nuclei~\cite{Voloshin:2010ut}, where the relative contributions of the
background (proportional to the elliptic flow) and the CME
(proportional to the magnetic field), should be very different in the
tip-tip and body-body type collisions. The second possibility would be
to exploit the large flow fluctuations in heavy-ion collisions as
discussed in~\cite{Voloshin:2010ut,Bzdak:2011my} and the ESE would be
a technique to perform such an analysis. (Note also that the magnetic
field depends very weakly on the initial shape geometry
fluctuations~\cite{Bzdak:2011my}.)  Yet another test, proposed
in~\cite{Voloshin:2011mx}, is based on the idea that the CME, the
charge separation along the magnetic field, should be zero if measured
with respect to the 4-th harmonic event planes, while the background
effects due to flow should be still present, albeit smaller in
magnitude ($\sim v_4$).  An example of such a correlator, would be $
\mean{\cos(2\phi_\alpha+2\phi_\beta-4\Psi_4}$, where $\Psi_4$ is the
fourth harmonic event plane.  The value of the background due to flow
could be estimated by rescaling the correlator Eq.~\ref{eq:obs1}.
Such measurements will require good statistics, and {\em strong}
fourth harmonic flow. Again, the ESE can be very helpful to vary the
effects related to flow.

\vspace{1mm}
\noindent {\em Measuring the shape and freeze-out velocity profile with
  azimuthally sensitive femtoscopy}.  Different shapes in the initial
geometry of the collision, to a different degree will be preserved in
the system freeze-out shapes. It was shown in~\cite{Voloshin:2011mg}
that those shapes can be addressed experimentally with azimuthally
sensitive femtoscopic analysis~\cite{Voloshin:1995mc,Voloshin:1997jh},
which has a goal to obtain the geometry of the source relative to
different harmonic symmetry planes. 
Such an analysis would definitely profit from event with extreme values
of anisotropy provided by the ESE, as the variation of
  femtoscopic parameters with azimuth would be better pronounced.
General details of femtoscopic analyses and discussion of the experimental
results can be found in a review~\cite{Lisa:2005dd}.

\vspace{1mm}
\noindent {\it Summary.}
Event shape engineering, providing possibility to study events
corresponding to nuclear collisions with
different initial geometry configuration, promises wide use in further
studies of the properties of the strongly interacting matter. 

\section{acknowledgements}
The authors are indebted to our colleagues at the ALICE Collaboration 
for numerous fruitful discussion.

\end{document}